\documentclass[aps,prd,twocolumn]{revtex4-1}
\makeatletter

\newcommand{\Rmnum}[1]{\expandafter\@slowromancap\romannumeral #1@}
\makeatother
\usepackage{epsfig}
\usepackage{mathrsfs}
\usepackage{slashed}
\usepackage{amsmath}
\usepackage{verbatim}
\usepackage{graphicx}
\usepackage{amssymb}
\usepackage{psfrag}
\usepackage{caption}
\newcommand{\ud}{\mathrm{d}}
\newcommand{\ue}{\mathrm{e}}
\newcommand{\ui}{\mathrm{i}}
\newcommand{\ug}{\mathrm{g}}

\newcommand{\bp}{\boldsymbol{p}}

\newcommand{\btau}{\boldsymbol{\tau}}

\newcommand{\bcdot}{\boldsymbol{\cdot}}
\newcommand{\MeV}{\mathrm{MeV}}

\begin{document}
\author{Yi-lun Du,$^{1}$ Zhu-fang Cui$^{2,3}$, Yong-hui Xia$^{2,3}$ and Hong-shi Zong$^{2,3,4}$}\email[]{zonghs@chenwang.nju.edu.cn}
\address{$^{1}$ Key Laboratory of Modern Acoustics, MOE, Institute of Acoustics, and Department of Physics, Nanjing University, Nanjing 210093, China}
\address{$^{2}$ Department of Physics, Nanjing University, Nanjing 210093, China}
\address{$^{3}$ State Key Laboratory of Theoretical Physics, Institute of Theoretical Physics, CAS, Beijing 100190, China}
\address{$^{4}$ Joint Center for Particle, Nuclear Physics and Cosmology, Nanjing 210093, China}
\title{Discussions on the crossover property within the Nambu--Jona-Lasinio model}
\begin{abstract}
In this paper, chiral symmetry breaking and its restoration are investigated in the mean field approximation of Nambu--Jona-Lasinio model. A first-order phase transition exists at low temperature, but is smeared out at high temperature. We discuss the rationality of using susceptibilities as the criteria to determine the crossover region as well as the critical point. Based on our results, it is found that to define a critical band instead of an exclusive line in this region might be a more suitable choice.

\bigskip

\noindent PACS Number(s): 12.39.-x, 25.75.Nq, 12.39.Fe
\end{abstract}
\maketitle

\section{Introduction}
Quantum chromodynamics (QCD)~\cite{PhysRevLett.30.1343,PhysRevLett.30.1346} is often viewed as the basic theory of strong interaction. In the process of large momentum transfer, the coupling constant is small (which is the so-called asymptotic freedom phenomenon~\cite{DavidPolitzer1974129}), so that the scattering process can be treated perturbatively with much success. However, in the
process of small momentum transfer, the coupling constant becomes large so that problems have to be treated with various nonperturbative methods.

The Nambu--Jona-Lasinio model (NJL)~\cite{PhysRev.122.345,PhysRev.124.246} is an effective model of QCD which was proposed in 1961. In this model, interaction terms are treated as four-body interactions, meanwhile, the Lagrangian is constructed such that the basic symmetries of QCD that are observed in nature are part and parcel of it. In particular, the NJL model exhibits the feature of dynamical chiral symmetry breaking~\cite{PhysRevLett.4.380,goldstone1961field}, which is responsible for the dynamical mass generation from bare quarks. Nevertheless, there are also two shortcomings of the NJL model; namely, it is neither confining nor renormalizable. As for the former one, the NJL model is expected to work well in the region of intermediate length between the asymptotic freedom and confinement regions and applied to the properties for which confinement is expected not to be essential. For the latter shortcoming, a momentum cutoff is often
introduced to avoid the ultraviolet divergence.

At high temperature and/or high density, the features of confinement and chiral symmetry breaking are expected to be destroyed. The effective quark mass and the pion mass may change discontinuously at certain point of temperature and density, which corresponds to the chiral phase transition. In the chiral limit, the quark condensate can be regarded as an order parameter of this phase transition, however, how to define the quark condensate beyond the chiral limit from first principles of QCD is still an open problem~\cite{maris1998pion}, and so there is no rigorous order parameter right now. Instead, the effective quark mass $M$ (or the quark condensate beyond chiral limit, although it has no rigorous definition) and various susceptibilities are often used as the criteria to determine the critical point of chiral phase transition~\cite{Masayuki1989668,he2009crossover,shi2010calculation,jiang2011model,li2013calculation}. In this work, we try to study the critical point of phase transition in the case of finite temperature and finite chemical potential by means of several susceptibilities in the NJL model.

The rest of this paper is organized as follows. In Sec. \Rmnum{2} the NJL model is briefly reviewed, and we will treat problems in the mean field approximation of this model. In Sec. \Rmnum{3} we present the results of M.Asakawa and K.Yazaki~\cite{Masayuki1989668} on the chiral phase transition, and question the rationality of their criterion for the critical point in the crossover region. In Sec.
\Rmnum{4}, several susceptibilities, which are expected to be the criteria for the critical point, are calculated. Finally, in Sec. \Rmnum{5} we will summarize our results and give the conclusions.

\section{The Nambu--Jona-Lasinio Model}
As an important effective theory, NJL model is widely used in many fields, for example, Refs \cite{zhao2008chiral,andersen2010pion,PhysRevC.82.015809,PhysRevD.81.125010,PhysRevLett.106.142003,PhysRevD.86.074018,
PhysRevD.87.054004,cui2013wigner} are some recent representative applications. The Lagrangian of the NJL model~\cite{Masayuki1989668} is (in this paper, we take the number of flavors $N_f=2$, and the number of colors $N_c=3$.)
\begin{equation}
\mathscr{L}_{\text{NJL}} =\bar{\psi}({\ui}\slashed{\partial}-m)\psi+{\ug}
                 [(\bar{\psi}\psi)^2+(\bar{\psi}{\ui}\gamma_5{\btau}
                 \psi)^2],
\end{equation}
where $m$ is the current quark mass for two flavors, $g$ is a coupling constant with the dimension of mass$^{-2}$, and the flavor and color indices are suppressed.

The thermal expectation value of an operator $\Theta$ is denoted as
\begin{equation}
\langle\!\langle\Theta\rangle\!\rangle=\frac{\mathrm{Tr}\,\Theta\,{\ue}^{-\beta(\mathscr{H}-\mu\mathscr{N}})}{\mathrm{Tr}\,
{\ue}^{-\beta(\mathscr{H}-\mu\mathscr{N})}}.
\end{equation}
Then, we apply the mean field approximation~\cite{kunihiro1984self,hatsuda1985soft}
to $\bar{\psi}\psi,\bar{\psi}\gamma_0\psi$ in the original and
Fierz-transformed interaction terms, while other terms vanish. Hence the mean field interaction terms can be written as
\begin{equation}\label{eqn:lmfint}
\begin{aligned}
\mathscr{L}_{\text{MFint}}=&\frac{4N_c+1}{2N_c}{\ug}\langle\!\langle\bar{\psi}\psi\rangle\!\rangle\bar{\psi}\psi+\frac{\ug}{N_c}
\langle\!\langle\bar{\psi}\gamma_0\psi\rangle\!\rangle\bar{\psi}\gamma_0\psi\\
&-\frac{4N_c+1}{4N_c}{\ug}\langle\!\langle\bar{\psi}\psi\rangle\!\rangle^2+\frac{\ug}{2N_c}\langle\!\langle\bar{\psi}\gamma_0
\psi\rangle\!\rangle^2,
\end{aligned}
\end{equation}
and the Hamiltonian density is
\begin{equation}\label{eqn:h}
  \begin{aligned}
  \mathscr{H}= & -{\ui}\bar{\psi}{\boldsymbol
           {\gamma}}{\bcdot}{\boldsymbol{\nabla}}\psi+m\bar{\psi}\psi-\mathscr{L}_{\text{MFint}}\\
             = & -{\ui}\bar{\psi}{\boldsymbol {\gamma}}{\bcdot}{\boldsymbol{\nabla}}\psi+M\bar{\psi}\psi\\
               &+\frac{\ug}{N_c}\sigma_2\mathscr{N}+G\sigma_1^2-\frac{\ug}{2N_c}\sigma_2^2,
   \end{aligned}
\end{equation}
where $G=\frac{4N_c+1}{4N_c}{\ug}$ is the renormalized coupling constant, $\mathscr{N}=\bar{\psi}\gamma_0\psi=\psi^\dagger\psi$ is the operator for the quark number density and $M$ is the effective quark mass
\begin{equation}\label{eqn:M}
  M=m-2G\sigma_1,
\end{equation}
$\sigma_1$ and $\sigma_2$ are defined, respectively, as follows:
\begin{equation}\label{eqn:sigma1}
  \sigma_1=\langle\!\langle\bar{\psi}\psi\rangle\!\rangle,
\end{equation}
\begin{equation}\label{eqn:sigma2}
  \sigma_2=\langle\!\langle\psi^\dagger\psi\rangle\!\rangle,
\end{equation}
where $\sigma_1$ is the quark condensate which is often viewed as the order parameter for chiral phase transition in the chiral limit.

The Hamiltonian density in (\ref{eqn:h}) describes a system of free quarks with mass $M$ and  chemical potential $\mu_r$ given by
\begin{equation}\label{eqn:mur}
  \mu_r=\mu-\frac{{\ug}}{N_c}\sigma_2,
\end{equation}
where $\mu$ is the bare chemical potential.

Now we use the formalism of the thermal Green function in the real time~\cite{PhysRevD.9.3320} to determine $\sigma_1$ and $\sigma_2$ self-consistently at finite temperature and chemical potential. The thermal Green function of a free fermion at temperature $T$ and chemical potential $\mu$ written in momentum space is as follows:
\begin{equation}\label{eqn:G1}
\begin{aligned}
 \hspace{-3mm}G(p;T,\mu)=&(\slashed{p}+M)\bigg[\frac{1}{p^2-M^2+{\ui}\varepsilon}+2\pi{\ui}\delta(p^2-M^2)\\
&\times\{\theta(p^0)n(\boldsymbol{p},\mu)+\theta(-p^0)m(\boldsymbol{p},\mu)\}\bigg],
\end{aligned}
\end{equation}
with
\begin{equation}\label{eqn:nm}
  n(\boldsymbol{p},\mu)=\frac{1}{1+\exp{[\beta(E-\mu)]}},
\end{equation}
\begin{equation}
  m(\boldsymbol{p},\mu)=\frac{1}{1+\exp{[\beta(E+\mu)]}},
\end{equation}
$E_p=\sqrt{M^2+{\bp}^2}$ and $\beta=1/T$. Using this propagator,
 $\sigma_1$ and $\sigma_2$ can be calculated out~\cite{Masayuki1989668}. The results are written
as
 \begin{gather}
 \begin{split}
 \hspace{-2mm}\sigma_1=-M\frac{N_cN_F}{\pi^2}\!\int_0^\Lambda\!
  \frac{p^2}{E}\big\{1-n(\boldsymbol{p},\mu_r)-
  m(\boldsymbol{p},\mu_r)\big\}{\ud}p\label{eq:sigma11}\\
  \end{split}\\
\begin{split}
 \hspace{-20mm} \sigma_2=\frac{N_cN_F}{\pi^2}\int_0^\Lambda
  p^2\big\{n(\boldsymbol{p},\mu_r)-m(\boldsymbol{p},\mu_r)\big\}{\ud}p, \label{eq:sigma22}
\end{split}
 \end{gather}
where $\Lambda$ is a momentum cutoff, which is introduced to avoid the ultraviolet divergence mentioned above. For convenience we will use the noncovariant three-momentum cutoff
scheme ~\cite{PhysRevD.34.1601}.

Now Eqs. (\ref{eqn:M}),(\ref{eqn:mur}),(\ref{eq:sigma11}) and (\ref{eq:sigma22}) form a set of self-consistent equations. By solving these equations self-consistently, one can obtain the effective quark mass for each temperature and chemical potential. In this paper, we will employ the widely accepted parameter set according to Hatsuda and Kunihiro~\cite{Hatsuda1994221}: $m=5.5~{\MeV}, \Lambda=631~{\MeV},
{\ug}=5.074\times10^{-6}~{\MeV}^{-2}$, which yields pion mass $M_\pi=138~{\MeV}$, pion decay constant $f_\pi=93.1~{\MeV}$, and quark condensate ${\langle\!\langle\bar{\psi}\psi\rangle\!\rangle}^{1/3}=-331~{\MeV}$.

\section{chiral restoration and the phase diagram}
\begin{figure}[bp]
  \centering
  \includegraphics[width=7.0cm]{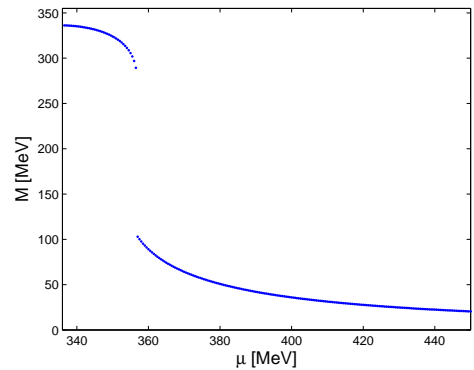}
  \caption{(color online). The effective quark mass $M$ at $T$=0}
  \label{fig:gra1}
\end{figure}

In the case of zero temperature and finite chemical potential, the effective quark mass $M$ obtained  numerically using the iterative method is shown in Fig. \ref{fig:gra1}.
\captionsetup{font=small,labelsep=none}

As can be seen from Fig. \ref{fig:gra1}, there is a discontinuity of $M$ at some certain chemical potential, the behavior of the quark condensate $\sigma_1$ is the same as that of $M$ [see Eq. (\ref{eqn:M})], we can conclude that if $M$ can serve as an order parameter in this case, a first-order phase transition occurs at a coexistence chemical potential and zero temperature. One defect, although it does not affect our subsequent discussions on the crossover property and the quantitative results in the first order transition, should be pointed out. As many literatures, e.g.,~\cite{sasaki2008chiral,pinto2012surface} show, the gap equation has more than one solution including an unstable solution and a metastable solution in the vicinity of the coexistence chemical potential in the first-order phase transition region. In order to select the solutions corresponding to stable points, it is wiser to find the smaller minima of the effective thermodynamical potential. The phase transition line can be located at the chemical potential where the effective thermodynamical potential has two degenerate minima. Then, with the increase of temperature, the two degenerate minima of the effective thermodynamical potential will get closer, i.e., the discontinuity of $M$ becomes small, and at last the discontinuity of $M$ comes to disappear at a critical temperature $T_c$, which means the first-order phase transition is smeared out. It is worthwhile to mention that the equations which the critical temperature $T_c$ should satisfy can be obtained by other techniques such as Landau's expansion or the method presented by Contrera \textit{et al.} ~\cite{contrera2010nonlocal}. Avancini \textit{et al.}. also provide other alternative to determine the first order line, the critical endpoint as well as the crossover region~\cite{avancini2012qcd}.

\begin{figure}[!h]
  \centering
  \includegraphics[width=7.0cm]{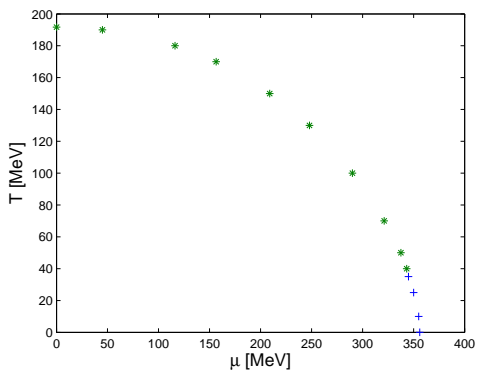}
  \caption{(color online). Phase transition line in the $T-\mu$ plane}
  \label{fig:gra3}
\end{figure}

At supercritical temperature $T>T_c$, there is no evident critical line which can be defined in the $T-\mu$ plane. However, the thermodynamical properties change rather sharply across a band in the plane, thus it is necessary to draw the phase diagram neglecting the smearing. M.Asakawa and K.Yazaki~\cite{Masayuki1989668} pointed out that there is a region where the effective quark mass $M$ changes very rapidly with temperature and chemical potential, and when the first-order phase transition is observed ($T<T_c$), $M$ jumps from a value larger than or around $\frac{1}{2}M_0$ to a value smaller than or around $\frac{1}{2}M_0$, where $M_0$ is denoted as the effective quark mass at $T=0$ and $\mu=0$. At $T=T_c$, $M$ changes most rapidly at the point where $M\simeq\frac{1}{2}M_0$. Considering that the critical line should be continuous at $T=T_c$, they artificially define the critical points as the points where the effective quark mass takes the same value as that at $T=T_c$, i.e., $\frac{1}{2}M_0$. According to this criterion, the phase transition diagram is plotted in Fig. \ref{fig:gra3}, where the ``+'' lines stand for the first-order phase transition lines and the ``*'' lines stand for the smooth transition regions as defined above.

Considering that the artificially defined critical point is aimed to describe the rapid change of thermodynamical properties, more convincing criteria for the critical point are the extremum of susceptibilities, such as the chiral susceptibility and the quark number susceptibility~\cite{jiang2011model,he2009crossover}. We will move on to discuss these in the next section of this paper.

\section{various susceptibilities and the crossover region}
Now let us introduce the definitions of four kinds of susceptibilities: the chiral susceptibility $\chi_s$, the quark number susceptibility $\chi_q$, the vector-scalar susceptibility $\chi_{vs}$, and another auxiliary susceptibility  $\chi_m$. For mathematical convenience we first introduce these susceptibilities in the free quark gas case (the interaction terms in the Lagrangian is zero, i. e., $\mathscr{L}_{int}=0$)~\cite{kunihiro2000chiral}, where $M$ and $\mu_r$ are reduced to $m$ and $\mu$, which are independent quantities. Denoting them with the superscript $^{(0)}$, their definitions and expressions are as follows:
\begin{gather}
\begin{split}
 \hspace{0mm}\chi_s^{(0)}\equiv&-\frac{\partial\langle\!\langle\bar{\psi}\psi\rangle\!\rangle_f}{\partial m}\\
  =&\frac{N_cN_f}{\pi^2}\int_0^\Lambda\bigg[\frac{m^2p^2\beta}{E^2}g(\mu)+\frac{p^4}{E^3}f(\mu)\bigg]{\ud}p, \label{eq:chis}
\end{split}\\
\begin{split}
 \hspace{-10mm}\chi_q^{(0)}\equiv&\frac{\partial\langle\!\langle\psi^{\dagger}\psi\rangle\!\rangle_f}{\partial \mu}=\frac{N_cN_f}{\pi^2}\int_0^\Lambda p^2\beta g(\mu){\ud}p, \label{eq:chiq}
\end{split}\\
\begin{split}
 \hspace{-10mm}\chi_{vs}^{(0)}\equiv&\frac{\partial\langle\!\langle\bar{\psi}\psi\rangle\!\rangle_f}{\partial \mu} =\frac{N_cN_f}{\pi^2}\int_0^\Lambda\frac{mp^2\beta}{E}h(\mu){\ud}p, \label{eq:chivs}
\end{split}\\
\begin{split}
 \hspace{-34mm}\chi_{m}^{(0)}\equiv-\frac{\partial\langle\!\langle\psi^{\dagger}\psi\rangle\!\rangle_f}{\partial m}=\chi_{vs}^{(0)}, \label{eq:chim}
\end{split}
\end{gather}
where $g(\mu)+h(\mu)=2n(\mu)(1-n(\mu))$, $g(\mu)-h(\mu)=2m(\mu)(1-m(\mu))$, $f(\mu)=1-n(\mu)-m(\mu)$,
the subscript $f$ represents the free quark gas systems. It should be noted that $\chi_{m}^{(0)}$ and $\chi_{vs}^{(0)}$ have the same analytical expression, which is reasonable from the viewpoint of statistical mechanics:
\begin{equation}\label{eqn:mixed partial derivative}
\begin{aligned}
\chi_{m}^{(0)}=\chi_{vs}^{(0)}=\frac{T}{V}\frac{\partial^2}{\partial m \partial \mu}\ln Z_{f},
\end{aligned}
\end{equation}
where $Z_{f}$ is the QCD partition function in the free quark gas case.

In the interacting case, $M$ and $\mu_r$ are no longer independent. These susceptibilities are coupled with each other as follows:
\begin{gather}
\begin{split}
 \hspace{0mm}\chi_s\equiv&-\frac{\partial\langle\!\langle\bar{\psi}\psi\rangle\!\rangle}{\partial m}\\
 =&\chi_{s}^{(0)}(\mu_r)(1+2G\chi_s)-\frac{{\ug}}{N_c}\chi_{vs}^{(0)}(\mu_r)\chi_m, \label{eq:chisc}
\end{split}\\
\begin{split}
 \hspace{0mm}\chi_q\equiv&\frac{\partial\langle\!\langle\psi^{\dagger}\psi\rangle\!\rangle}{\partial \mu}\\
 =&2G\chi_{vs}^{(0)}(\mu_r)\chi_{vs}+ \chi_q^{(0)}(\mu_r)(1-\frac{{\ug}}{N_c}\chi_q)
,\label{eq:chiqc}
\end{split}\\
\begin{split}
 \hspace{0mm}\chi_{vs}\equiv&\frac{\partial\langle\!\langle\bar{\psi}\psi\rangle\!\rangle}{\partial \mu}\\
 =& 2G\chi_{s}^{(0)}(\mu_r)\chi_{vs}+\chi_{vs}^{(0)}(\mu_r)(1-\frac{{\ug}}{N_c}\chi_q)
 ,\label{eq:chivsc}
\end{split}\\
\begin{split}
 \hspace{0mm}\chi_{m}\equiv&-\frac{\partial\langle\!\langle\psi^{\dagger}\psi\rangle\!\rangle}{\partial m}\\
 =&\chi_m^{(0)}(\mu_r)(1+2G\chi_s)-\frac{{\ug}}{N_c}\chi_q^{(0)}(\mu_r)
 \chi_m.\label{eq:chimc}
\end{split}
\end{gather}

\begin{figure}[t]
  \centering
  \includegraphics[width=7.0cm]{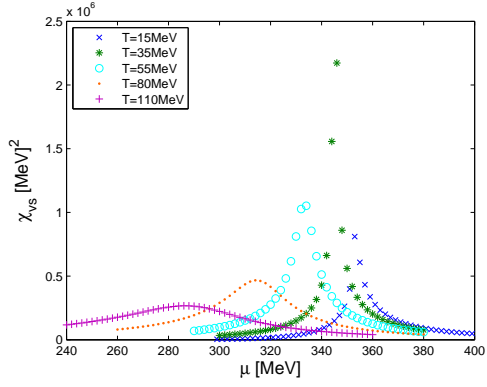}
  \caption{(color online). $\chi_{vs}$ at different $\mu$ and $T$}
  \label{fig:gra7chivsc}
\end{figure}
Using the iterative method, we can obtain the numerical results of these susceptibilities. For example, the vector-scalar susceptibility is the response of the effective quark mass ($M=m-2G\langle\!\langle\bar{\psi}\psi\rangle\!\rangle$) to the chemical potential $\mu$, and its result is shown in Fig. \ref{fig:gra7chivsc}.
\begin{figure}[!b]
  \centering
  \includegraphics[width=7.0cm]{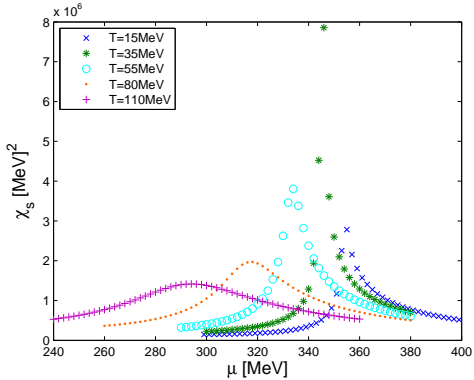}
  \caption{(color online). $\chi_{s}$ at different $\mu$ and $T$}
  \label{fig:gra7chisc}
\end{figure}
It can be seen that, when $T$ is smaller than the critical value $T_c=35$~MeV, there always exists a convergent discontinuity of $\chi_{vs}$, corresponding to a first-order phase transition; when $T=T_c$, $\chi_{vs}$ displays a sharp and narrow divergent peak, which implies a second-order phase transition, or in other words, here is a critical end point; when $T>T_c$, the discontinuity disappears and a rather broad peak of finite height is shown, corresponding to the crossover region. We pick the peak of the susceptibility as the artificial critical point to draw the phase diagram, which would produce little change on Fig. \ref{fig:gra3}.

\begin{figure}[t]
  \centering
  \includegraphics[width=7.0cm]{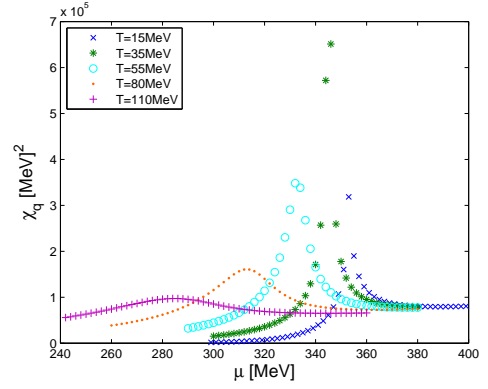}
  \caption{(color online). $\chi_{q}$ at different $\mu$ and $T$}
  \label{fig:gra7chiqc}
\end{figure}

Comparing these with the results of chiral susceptibility $\chi_s$ and quark number susceptibility $\chi_q$, shown in Fig. \ref{fig:gra7chisc} and \ref{fig:gra7chiqc}, respectively, we can conclude that at low temperature, the first-order phase transition occurs at almost the same chemical potential, while in the crossover region, the artificially defined critical point tends to occur at different chemical potentials as the temperature increases.

\begin{figure}[!b]
  \centering
  \includegraphics[width=7.0cm]{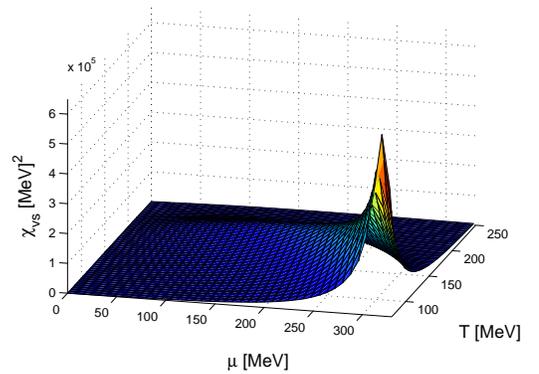}
  \caption{(color online). Vector-scalar susceptibility $\chi_{vs}$ in the crossover region}
  \label{fig:chivs3d}
\end{figure}

\begin{figure}[!t]
  \centering
  \includegraphics[width=7.0cm]{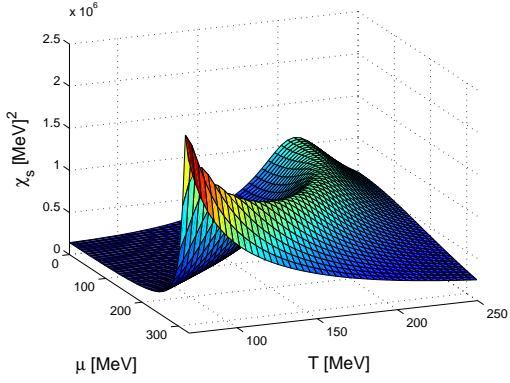}
  \caption{(color online). Chiral susceptibility $\chi_{s}$ in the crossover region}
  \label{fig:chis3d}
\end{figure}

\begin{figure}[!b]
  \centering
  \includegraphics[width=7.0cm]{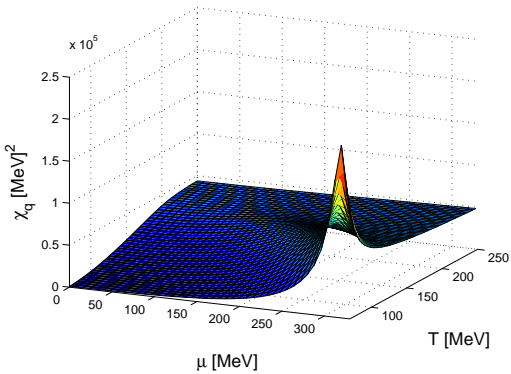}
  \caption{(color online). Quark number susceptibility $\chi_{q}$ in the crossover region}
  \label{fig:chiq3d}
\end{figure}

The calculated results of $\chi_{vs}$, $\chi_s$, and $\chi_q$ in the crossover region are shown in Figs. \ref{fig:chivs3d}-\ref{fig:chiq3d}, respectively. $\chi_{vs}$ and $\chi_{m}$ give the same numerical results for the reason mentioned above. We can find that they exhibit different behaviors: the chiral susceptibility $\chi_s$ exhibits an obvious band, so it is convincing to define the peak of $\chi_s$ as the artificial critical point; in the high $T$ and/or low $\mu$ region, the vector-scalar susceptibility $\chi_{vs}$ tends to vanish; while the global shape of the quark number susceptibility $\chi_q$ is just similar to the ones of $\chi_{s}$ and $\chi_{vs}$, but it is nonvanishing in the high $T$ and/or high $\mu$ region  whose behavior is closely linked to the quark number density. This result is consistent with K. Fukushima's result obtained using the PNJL model~\cite{fukushima2008phase}. Therefore, $\chi_q$ cannot describe the crossover property well in the high $T$ and/or high $\mu$ region.

It is very interesting and meaningful to compare the results of these susceptibilities with that of the thermal susceptibility $\chi_T=\partial\langle\!\langle\bar{\psi}\psi\rangle\!\rangle/\partial T$ which is widely used in many recent literatures, e.g.,~\cite{morita2011role,skokov2012phase}. For mathematical convenience, we define $\chi_n=\partial\langle\!\langle\psi^{\dagger}\psi\rangle\!\rangle/\partial T$. By the same process employed above, we obtain a set of coupled equations for $\chi_T$ and $\chi_n$ as follows:

\begin{gather}
\begin{split}
 \hspace{0mm}\chi_T\equiv&\frac{\partial\langle\!\langle\bar{\psi}\psi\rangle\!\rangle}{\partial T}\\
 &=\frac{\partial\langle\!\langle\bar{\psi}\psi\rangle\!\rangle}{\partial M}\frac{\partial M}{\partial T}+\frac{\partial\langle\!\langle\bar{\psi}\psi\rangle\!\rangle}{\partial \mu_r}\frac{\partial \mu_r}{\partial T}+\big(\frac{\partial\langle\!\langle\bar{\psi}\psi\rangle\!\rangle}{\partial T}\big)_{M,\mu_r}\\
 =&2G\chi_s^{(0)}(\mu_r)\chi_T-\frac{{\ug}}{N_c}\chi_{vs}^{(0)}(\mu_r)\chi_n\\
 &+M\beta\chi_q^{(0)}(\mu_r)-\mu_r\beta\chi_m^{(0)}(\mu_r), \label{eq:chiT}
\end{split}\\
\begin{split}
 \hspace{0mm}\chi_n\equiv&\frac{\partial\langle\!\langle\psi^{\dagger}\psi\rangle\!\rangle}{\partial T}\\
=&\frac{\partial\langle\!\langle\psi^{\dagger}\psi\rangle\!\rangle}{\partial M}\frac{\partial M}{\partial T}+\frac{\partial\langle\!\langle\psi^{\dagger}\psi\rangle\!\rangle}{\partial \mu_r}\frac{\partial \mu_r}{\partial T}+\big(\frac{\partial\langle\!\langle\psi^{\dagger}\psi\rangle\!\rangle}{\partial T}\big)_{M,\mu_r}\\
 =&2G\chi_{m}^{(0)}(\mu_r)\chi_{T}- \frac{{\ug}}{N_c}\chi_q^{(0)}(\mu_r)\chi_n\\
 &-\mu_r\beta\chi_q^{(0)}(\mu_r)+\frac{N_cN_f}{\pi^2}\int_0^\Lambda p^2E\beta^2h(\mu){\ud}p.\label{eq:chin}
\end{split}
\end{gather}

\begin{figure}[!t]
  \centering
  \includegraphics[width=7.0cm]{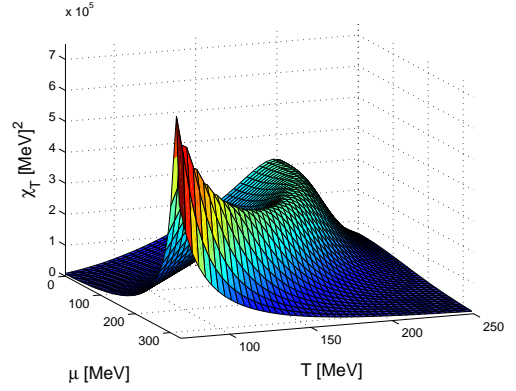}
  \caption{(color online). Thermal susceptibility $\chi_{T}$ in the crossover region}
  \label{fig:chiT3d}
\end{figure}

\begin{figure}[!b]
  \centering
  \includegraphics[width=7.0cm]{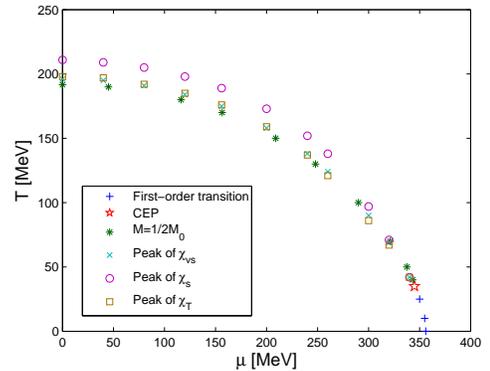}
  \caption{(color online).  Phase diagram obtained according to different criteria}
  \label{fig:phasedia}
\end{figure}

The behavior of $\chi_T$ in the crossover region is shown in Fig.\ref{fig:chiT3d}, which is very similar to that of $\chi_s$. The comparison between them will be shown in the phase diagram.

The phase diagram is given in Fig. \ref{fig:phasedia} according to different criteria for the critical point in the crossover region as follows: the peaks of $\chi_{vs}$, $\chi_T$, and $\chi_s$. We compare our results with that of M.Asakawa and K.Yazaki. The line given by the peak of $\chi_{vs}$ is almost the same as that of M.Asakawa and K.Yazaki, while the lines given by the peaks of $\chi_T$ and $\chi_s$ both display a shift as compared with that of M.Asakawa and K.Yazaki. In the $\mu=0$ case, the peak of $\chi_{vs}$ appears at $T=196$ MeV, while the peaks of $\chi_T$ and $\chi_s$ appear at $T=198$ MeV and $T=211$ MeV, respectively. Here we do not give the results of $\chi_q$ due to the reason mentioned above. Therefore, it is hard to define the critical line in the crossover region, and a critical band might be a more suitable choice.

\section{summary and conclusion}
In this paper we study the chiral phase transition at finite temperature and chemical potential and calculate several susceptibilities in the mean field approximation using the Nambu--Jona-Lasinio model.

We discuss the rationality of using susceptibilities as the criteria to determine the crossover region as well as the critical point. In the low temperature region, the first-order phase transition is found to be at almost the same chemical potential for different susceptibilities, which is due to the fact that these susceptibilities are coupled with each other in their mathematical form; at sufficiently high temperature, the first-order phase transition is smeared out, and the results of different susceptibilities imply the uncertainty in the position of the artificially defined critical point in the $T-\mu$ plane. So, it is more suitable to define a critical band rather than an exclusive line in the crossover region.

\acknowledgments
This work is supported in part by the National Natural Science Foundation of China (under Grant Nos. 11275097, 10935001, 11274166 and 11075075), the National Basic Research Program of China (under Grant No. 2012CB921504) and the Research Fund for the Doctoral Program of Higher Education (under Grant No. 2012009111002).

\bibliographystyle{apsrev4-1}
\bibliography{DJ11475}

\end{document}